\documentclass[twocolumn,runningheads]{svjour2}
\smartqed  % flush right qed marks, e.g. at end of proof
\usepackage{graphicx}
\journalname{Astrophysics and Space Science}

\newcommand{\be}{\begin{equation}}
\newcommand{\ee}{\end{equation}}
\newcommand{\ba}{\begin{eqnarray}}
\newcommand{\ea}{\end{eqnarray}}
\newcommand{\bc}{\begin{center}}
\newcommand{\ec}{\end{center}}

\begin{document}

\title{Collective effects of stellar winds and unidentified gamma-ray sources
}
\subtitle{}

\titlerunning{Stellar winds and unidentified gamma-ray sources}        % if too long for running head

\author{ Diego F. Torres   \& Eva Domingo-Santamar\'ia
        }


\institute{ DFT: Instituci\'o de Recerca i Estudis Avan\c{c}ats (ICREA)
\&        Institut de Ci\`encies de l'Espai (IEEC-CSIC),
              Facultat de Ciencies, 
              Universitat Aut\`onoma de Barcelona,
              Torre C5 Parell, 2a planta, 08193 Barcelona, Spain
               \email{dtorres@ieec.uab.es}. \at EDS:
Institut de F\'{\i}sica d'Altes Energies (IFAE), Edifici C-n, Campus
UAB, 08193 Bellaterra, Spain}

\date{Received: date / Accepted: date}

\maketitle

\begin{abstract}
We study collective wind configurations produced by a number of
massive stars, and obtain densities and expansion velocities of the
stellar wind gas that is to be target, in this model, of hadronic
interactions. We study the expected $\gamma$-ray emission from these
regions, considering in an approximate way the effect of cosmic ray
modulation. We compute secondary particle production (electrons from
knock-on interactions and electrons and positrons from charged pion
decay), and solve the loss equation with ionization, synchrotron,
bremsstrahlung, inverse Compton, and expansion losses. We provide
examples where configurations can produce sources for GLAST
satellite, and the MAGIC, HESS, or VERITAS telescopes in non-uniform
ways, i.e., with or without the corresponding counterparts. We show
that in all cases we studied no EGRET source is expected. 

\keywords{$\gamma$-rays \and unidentified $\gamma$-ray sources}
\PACS{}
\end{abstract}

\section{Single and collective stellar winds}

LBL, WR, O and B stars lose a significant fraction of their mass in stellar
winds with terminal velocities that can easily reach $V_\infty \sim 10^3$ km
s$^{-1}$. With mass loss rates as high as
$\dot{M}_\star=(10^{-6}-10^{-4})$ M$_\odot$ yr$^{-1}$, the
density at the base of the wind can reach $10^{-12}$ g cm$^{-3}$
(e.g., \cite{1}). Such winds are
permeated by significant magnetic fields, and provide a matter
field dense enough as to produce hadronic $\gamma$-ray emission
when pervaded by relativistic particles. A typical wind
configuration (\cite{1,2,3}) contains an inner
region in free expansion (zone I) and a much larger hot compressed
wind (zone II). These are finally surrounded by a thin layer of
dense swept-up gas (zone III); the final interface with the
interstellar medium (ISM). The innermost region size is fixed by
requiring that at the end of the free expansion phase (about 100
years after the wind turns on) the swept-up material is comparable
to the mass in the driven wave from the wind, which happens at a
radius $R_{\rm wind}=V_\infty (3\dot{M}_\star/4 \pi \rho_0
V_\infty^3)^{1/2}$, where $\rho_0\approx m_p n_0$ is the ISM mass
density, with $m_p$ the mass of the proton and $n_0$ the ISM
number density. After hundreds of thousands of years, the wind
produces a bubble with a radius of the order of tens of parsecs,
with a density lower (except that in zone I) than in the ISM.
The matter in the inner region is described through the
continuity equation: $\dot{M}_\star=4\pi r^2 \rho (r) V(r)$, where
$\rho(r)$ is the density of the wind and $V(r)=V_{\infty} (1-{R_0}/{r})^{\beta}$ is its velocity.
%Hence,
%\begin{equation}
%\rho(r)=\frac{\dot{M}_\star}{4\pi r^2 v(r)}.
%\end{equation}
%The latter is $ V(r)=V_{\infty} (1-{R_0}/{r})^{\beta}, $ where
$V_{\infty}$ is the terminal wind velocity, and the parameter
$\beta$ is $\sim 1$ for massive stars (\cite{1}). $R_0$ is given in terms of the wind velocity close to the
star, $V_0 \sim 10^{-2}V_\infty$, as $R_0=R_\star (1-(V_0/V_\infty)^{1/\beta})$.
Hence, the particle density
is $n(r)=\dot {M}_\star (1-{R_0}/{r} )^{-\beta}/({4\pi m_p
V_{\infty} r^2}). $ In the case of a collection of winds, a hydrodynamical model
taking into account the composition of different single stellar winds needs to be adopted. As we see below, differences with the single stellar wind case are notable.

Here we adopt a similar modelling to that of Cant\'o et al. (2000) \cite{3} (see
also \cite{5,6,7}) to describe the wind of a cluster (or a
sub-cluster) of stars.  Consider that there are $N$
stars in close proximity, uniformly distributed within a radius
$R_c$, with a number density
\be \label{n} n = \frac{3N}{ 4\pi R_c^3} \,. \ee
Each star has its own mass-loss rate ($\dot M_{i}$) and (terminal)
wind velocity ($V_i$), and average values can be defined as
\ba \dot M_w &=& \frac{1}{N} \sum_{i}^{N} \dot M_i \,,  \\
V_w &=& \left( \frac{\sum_{i}^{N} \dot M_i {V_i}^2}{N\,\dot M_w}
\right)^{1/2} \,. \ea
All stellar winds are assumed to mix with the surrounding ISM and
with each other, filling the intra-cluster volume with a hot,
shocked, collective stellar wind. A stationary flow in which mass
and energy provided by stellar winds escape through the outer
boundary of the cluster is established.
For an arbitrary distance $R$ from the center of the cluster, mass
and energy conservation imply that
\ba \frac{4 \pi}{3} R^3 n \dot M_w &=& 4 \pi R^2 \rho V \,, \\
\frac{4 \pi}{3} R^3 n \dot M_w \left( \frac 12 {V_w}^2 \right)&=&
4 \pi R^2 \rho V \left( \frac 12 {V}^2 + h\right) \,, \ea
where $\rho$ and $V$ are the mean density and velocity of the cluster
wind flow at position $R$ and $h$ is its specific enthalpy (sum of
internal energy plus the pressure times the volume),
\be h = \frac{\gamma}{\gamma-1} \frac{P}{\rho} \,, \ee
with $P$ being the mean pressure of the wind and $\gamma$ being
the adiabatic index (hereafter $\gamma=5/3$ to fix numerical
values). From the mass conservation equation we obtain
\be \label{mm} \rho V = \frac {n \dot M_w}{3}R \,, \ee
and the ratio of the two conservation equations imply
\be \label{fff} \frac 12 V^2 + h = \frac 12 {V_w}^2 \,. \ee
The equation of motion of the flow is
\be \label{motion} \rho V \frac {dV}{dR} =
-\frac{dP}{dR} - n \dot M_w V \,, \ee
which, introducing the adiabatic sound speed $c$,
\be \label{cc} c^2 = \gamma \frac {P}{\rho} \,, \ee
can be written as
\be \label{ddd} \rho V \frac {dV}{dR} = - \frac 1\gamma
\frac{d(\rho c^2)}{dR} - n \dot M_w V \,. \ee
From the definition of enthalpy and Equation (\ref{fff}),
the adiabatic sound speed can be expressed as
\be \label{ccc} c^2 = \frac {\gamma-1}{2} ({V_w}^2-V^2) \,. \ee
Using Equation (\ref{mm}), its derivative $d\rho$ and Equation
(\ref{ccc}) in (\ref{ddd}) one obtains
\be \frac {dR}{R} = \frac {dV}{V} \left[
\frac{(\gamma-1){V_w}^2-(\gamma+1)V^2}
{(\gamma-1) {V_w}^2 + (5\gamma+1)V^2}\right] \,, \ee
which can be integrated and expressed in more convenient
dimensionless variables ($v\equiv V/V_w$ and $r \equiv R/R_c$) as
follows
\be v \left[ 1+\frac {5\gamma+1}{\gamma-1}
v^2\right]^{-(3\gamma+1)/(5\gamma+1)} = Ar \,, \label{inside} \ee
with $A$ an integration constant.
When $R>R_c$, i.e., outside the cluster, by definition $n=0$ and the mass conservation equation is
\be \dot M_{\rm assoc} \equiv \frac{4\pi}{3} {R_c}^3 n \dot M_w =
4\pi R^2 \rho V \,, \label{eee} \ee
where the middle equality gives account of the contribution of all
stars in the association, and $\dot M_{\rm assoc} = \sum_i \dot
{M_i}$ is the mass-loss
rate at the outer boundary $R_c$.
Substituting Equation (\ref{ccc}) and (\ref{eee}) into the $n=0$
realization of Equation (\ref{ddd}) one obtains
\be -\frac {dR}{R} = \frac {dV}{V} \left[
\frac{(\gamma-1){V_w}^2-(\gamma+1)V^2} {2 (\gamma-1) (V_w^2 -
V^2)}\right] \,, \ee
and integrating, the velocity in this outside region is implicitly
defined from
\be v (1-v^2)^{1/(\gamma-1)} = B r^{-2}, \label{outside} \ee
with $B$ an integration constant.
Having constants $A$ and $B$ in Equations (\ref{inside}) and
(\ref{outside}), see below, the velocity at any distance from the
association center can be determined by numerically solving its
implicit definitions, and hence the density is also determined,
through Equation (\ref{mm}) or (\ref{eee}).

From Equation (\ref{outside}), two asymptotic branches can be
found. When $r \rightarrow \infty$, either $v  \rightarrow 0 $
(asymptotically subsonic flow) or $v  \rightarrow 1 $
(asymptotically supersonic flow) are possible solutions. The first
one (subsonic) produces the following limits for the density, the
sound speed and the pressure
\ba
\rho_\infty &=& \frac{\dot M_{\rm assoc}}{4\pi B {R_c}^2 V_w} \,, \\
c_\infty^2 &=& \frac{\gamma-1}{2} {V_w}^2 \,, \\
P_\infty &=& \frac{\gamma-1}{2 \gamma} \frac{\dot M_{\rm assoc}V_w}
{4\pi B {R_c}^2 } \,. \ea
%
%For the supersonic solution,
%
%\be \rho_\infty = c_\infty^2 = P_\infty = 0 \,. \ee
%
If $P_\infty$ is the ISM pressure far from the
association, the constant $B$ can be obtained as
\be \label{pin} B = \frac{\gamma-1}{2 \gamma}
\frac{\dot M_{\rm assoc}V_w}{4\pi P_\infty {R_c}^2 } \,. \ee

The velocity of the flow at the outer radius $r=1$ follows from
Equation (\ref{outside})
\be v_{r=1} (1-{v_{r=1}}^2)^{1/(\gamma-1)} = B \,, \label{v1} \ee
and continuity implies that
\be v_{r=1} \left[ 1+\frac {5\gamma+1}{\gamma-1}
{v_{r=1}}^2\right]^{-(3\gamma+1)/(5\gamma+1)}=A \,. \label{cont} \ee
Equation (\ref{inside}) implicitly contains the dependence of $v$
with $r$ in the inner region of the collective wind. Its left hand
side is an ever increasing function. Thus, for the equality to be
fulfilled for all values of radius (0$<r<$1), the right hand side
of the equation must reach its maximum value at $r$=1. Deriving
the right hand side of Eq. ({\ref{inside}), one can find the
velocity that makes it  maximum
\be v_{\rm max}= \left( \frac{\gamma-1}{\gamma+1 }\right)^{1/2} \,.
\label{max} \ee
Since $v$ grows in the inner region, the maximum velocity is
reached at $r=1$, and from Equation (\ref{v1}),
\be B= \left( \frac{\gamma-1}{\gamma+1 }\right)^{1/2}
\left( \frac 2{\gamma+1}\right)^{1/(\gamma-1)} .\ee
Continuity (Equation \ref{cont}) implies that the value of $A$ is
\be A= \left( \frac{
\gamma-1}{\gamma+1 }\right)^{1/2} \left( \frac
{\gamma+1}{6\gamma+2}\right)^{(3\gamma+1)/(5\gamma+1)}. \ee
With the former value of $B$, and from Equation (\ref{pin}), if
\be P_\infty < \frac 1\gamma \left( \frac{ \gamma-1}{\gamma+1
}\right)^{1/2} \left( \frac {\gamma+1}{2}
\right)^{\gamma/(\gamma-1)} \frac{\dot M_{\rm assoc} V_w}{4\pi
R_c^2} \,, \ee
the subsonic solution is not attainable (continuity of the
velocity flow is impossible) and the supersonic branch is the only
physically viable. In this regime, the flow leaves the boundary of
the cluster $R_c$ at the local sound speed $v_{\rm max}$ (equal to
1/2 for $\gamma=5/3$) and is accelerated until $v=1$ for $r
\rightarrow \infty$.

Examples of the supersonic flow
(velocity and particle density) for a group of stars generating
different values of $\dot M_{\rm assoc}$, $V_w$, and $R_c$, are
given in Table 1. The total mass contained up to 10 $R_c$ is also
included in the Table as an example. A typical configuration of a group of tens
of stars  may generate a wind in expansion with a
velocity of the order of 1000 km s$^{-1}$ and a mass between
tenths and a few solar masses within a few pc (tens of $R_c$). We
consider hadronic interactions with this matter.
\begin{table}[t]
\begin{center}
\caption{Examples of configurations of collective stellar winds.
The mass is that contained within 10 $R_c$, and is shown as an example. $n_0$ is
the central density.} \vspace{0.2cm}
\begin{tabular}{lccccc}
\hline
 Model    & $\dot M_{\rm assoc}$  & $V_w$          & $R_c$ & $n_0$& Wind mass\\
          & [M$_\odot$ yr$^{-1}$] &  [km s$^{-1}$] & pc    &  cm$^{-3}$    &[M$_\odot$ ]\\
\hline
    A & $10^{-4}$ & 800 & 0.1 & 210.0 & 0.13\\
    B & $10^{-4}$ & 800 & 0.3 & 23.3 & 0.39\\
    C & $5 \times 10^{-5}$ & 1000 & 0.2 & 20.9 & 0.11\\
    D & $2 \times 10^{-4}$ & 1500 & 0.4 & 13.9 & 0.56\\
    E & $2 \times 10^{-4}$ & 2500 & 0.2 & 33.5 & 0.17\\
\hline \hline
\end{tabular}
\end{center}
\label{cantotable}
\end{table}
However, it is to be noted that not all cosmic rays will be able to enter
the collective wind. The difference
between an {\it inactive target}, as that provided by matter in
the ISM, and an {\it active or expanding target}, as that provided
by matter in a single or a collective stellar wind, is indeed given by 
modulation effects.
The cosmic ray penetration into the jet outflow
depends on the parameter $\epsilon \sim V R/D$,
where $V$ is velocity of wind, 
and $D$ is the diffusion coefficient. $\epsilon$ measures the ratio
between the diffusive and the convective timescale of the
particles (e.g., \cite{8}).

In order to obtain an analytic expression for $\epsilon$ for a
particular star we consider that the diffusion coefficient within
the wind of a particular star is given by (\cite{3,8,9}) $ D \sim \frac 13 \lambda_r c ,
$ where $\lambda_r$ is the mean-free-path for diffusion in the
radial direction (towards the star). The use of the Bohm
parameterization seems justified,  contrary to the solar
heliosphere, since we expect that in the innermost region of a
single stellar wind there are many disturbances (relativistic
particles, acoustic waves, radiatively driven waves, etc.). In the
case of a collective wind, the collision of individual winds of
the particular stars forming the association also produce many
disturbances. 
The mean-free-path for scattering parallel to the magnetic field
($B$) direction is considered to be $\lambda_\| \sim 10 r_g = 10
E/eB$, where $r_g$ is the particle gyro-radius and $E$ its energy.
In the perpendicular direction $\lambda$ is shorter, $\lambda_\bot
\sim r_g$. The mean-free-path in the radial direction is then
given by $\lambda_r={\lambda_\bot}^2 \sin^2 \theta +
{\lambda_\|}^2 \cos^2 \theta = r_g ( 10 \cos^2 \theta + \sin^2
\theta)$, where $\cos^{-2} \theta = 1+(B_\phi/B_r)^2$. Here, the
geometry of the magnetic field for a single star is represented by
the magnetic rotator theory (\cite{10}; see also \cite{1,8}) \be
\frac{B_\phi}{B_r}=\frac{V_\star}{V_\infty}
\left(1+\frac{R}{R_\star} \right) \label{bp} \ee and \be
\label{br} B_r=B_\star \left(\frac{R_\star}{R}\right)^2 , \ee
where $V_\star$ is the rotational velocity at the surface of the
star, and $B_\star$ the surface magnetic field. Near the star the
magnetic field is approximately radial, while it becomes
tangential far from the star, where $\lambda_r$ is dominated by
diffusion perpendicular to the field lines. This approximation
leads ---when the distance to the star is large compared with that
in which the terminal velocity is reached, what happens at a few
stellar radii--- to values of magnetic field and diffusion
coefficient normally encountered in the ISM.

Using all previous formulae, \ba E^{\rm min}(r) \sim \frac{ 3 e
B_\star V_\infty (r-R_\star)}{ c} \left( \frac {R_\star}{
r}\right)^2 \times \hspace{1.8cm} \nonumber \\  \frac{\left (
1+\left( \frac { V_\star }{V_\infty} \left( 1+ \frac r{R_\star}
\right) \right)^2 \right) ^{3/2}  } {10+ \left( \frac{ V_\star
}{V_\infty} \left( 1+ \frac r{R_\star} \right) \right)^2 } .
\label{ep} \ea Equation (\ref{ep}) defines a minimum energy below
which the particles are convected away from the wind. $E^{\rm
min}(r)$ is an increasing function of $r$, the limiting value of
the previous expression being \ba E^{\rm min}(r \gg R_\star)
&\sim& \frac{ 3 e B_\star V_\infty R_\star}{ c} \left(
\frac{V_\star}{V_\infty} \right) \sim  4.3 \left(
\frac{B_\star}{10 {\rm G}}\right) \times \nonumber \\ & &  \left(\frac{V_\star}{0.1
V_\infty}\right) \left(\frac{R_\star}{12 {\rm R}_\odot}\right){\rm
TeV } \,.\ea
Therefore, particles that are not convected in the outer regions
are able to diffuse up to its base. Note that $E^{\rm min}(r \gg
R_\star)$ is a linear function of all $R_\star$, $B_\star$ and
$V_\star$, which is typically assumed as $ V_\star \sim 0.1
V_\infty$ (e.g., \cite{1}). There is a large
uncertainty in these parameters, about one order of magnitude. The
values of the magnetic field on the surface of O and B stars is
under debate. Despite deep searches, only $5$ stars were found to
be magnetic (with sizeable magnetic fields in the range of
$B_\star \sim 100$ G) (e.g.,  \cite{11} and references
therein) typical surface magnetic fields of OB stars are then
presumably smaller.

In the kind of collective wind, we consider that the collective
wind behaves as that of a single star having a radius equal to
$R_c$, and mass-loss rate equal to that of the whole association,
i.e., $\dot M_{\rm assoc}$. The wind velocity at $R_c$, $V_\star$
is given by Equation (\ref{max}). The order of magnitude of the
{\it surface } magnetic field (i.e., the field at $R=R_c$) is
assumed as the value corresponding to the normal decay of a single
star field located within $R_c$, for which a sensitive assumption
can be obtained using Equations (\ref{bp}) and (\ref{br}), ${\cal
O}(10^{-6}$) G. This results, for the whole association, in
\be [E^{\rm min}(r \gg R_\star)]^{\rm assoc} \sim 0.8 \left(
\frac{B(R_c)}{1 \mu{\rm G}}\right) \left(\frac{R_c}{0.1 {\rm
pc}}\right){\rm TeV } . \label{EminAssoc} \ee
The value of the magnetic field is close to that typical of the
ISM, and should be consider as an average (this kind of magnetic
fields magnitude was also used in modelling the unindentified
HEGRA source in Cygnus, \cite{12}). In
particular, if a given star is close to $R_c$ its contribution to
the overall magnetic field near its position will be larger, but
at the same time, its contribution to the opposite region (distant
from it 2 $R_c$) will be negligible. In what follows we consider
hadronic processes up to 10 -- 20 $R_c$, so that a value of the
magnetic field typical of ISM values is expected. We shall
consider two realizations of $[E^{\rm min}(r \gg R_\star)]^{\rm
assoc}$, 100 GeV and 1 TeV.

\section{$\gamma$-rays and secondary electrons from
       a cosmic ray spectrum with a low energy cutoff }

The pion produced $\gamma$-ray emissivity is obtained from the
neutral pion emissivity %$Q_{\pi^0}$
as { described in detail in the appendix of
\cite{13}. } For normalization purposes, we use the expression of the energy
density that is contained in cosmic rays, $\omega_{\rm CR} =
\int_E N(E)\, E\, dE$
%
%\be \omega_{\rm CR} = \int_E N(E)\, E \, dE \ee
%
and compare it to the energy contained in cosmic rays in the Earth
environment, $ \omega_{\rm CR,\oplus}(E)=\int_E N_{\oplus}(E)\,
E\, dE $, where $N_{p\,\oplus}$ is the  local cosmic ray
distribution obtained from the measured cosmic ray flux.
The Earth-like spectrum, $J_\oplus(E)$, is $2.2 E_{{\rm GeV}}^{-2.75}$ 
{\rm cm}$^{-2}$  {\rm s}$^{-1}$ {\rm sr}$^{-1}$ {\rm GeV}$^{-1} $ (e.g. \cite{12a,12b}), so that $ \omega_{\rm CR,\oplus} (E>1 {\rm
GeV}) \sim 1.5 \, {\rm eV cm^{-3}}$. This implicitly defines an
enhancement factor, $\varsigma$, as a function of energy \be
\varsigma(E)=\frac{\int_E N(E) \, E \, dE }{\omega_{\rm
CR,\oplus}(E)} . \label{K} \ee
We assume that $N(E)$ is a power law of the form $N(E) = K_p
E^{-\alpha}$. Values of enhancement $\gg 100$ at all energies are
typical of star forming environments (see, e.g., \cite{13,14,14b,17,18,19}) and they would ultimately depend on the spectral
slope of the cosmic ray spectrum and on the power of the
accelerator. For a fixed slope, harder than that found in the
Earth environment, the larger the energy, the larger the
enhancement, due to the steep decline ($\propto E^{-2.75}$) of the
local cosmic ray spectrum. In what follows, as an example, we
consider enhancements of the full cosmic ray spectrum (for
energies above 1 GeV) of 1000. With such fixed $\varsigma$, the
normalization of the cosmic ray spectrum, $K_p$, can be obtained
from Equation (\ref{K}) for all values of the slope. Note that
$K_p \propto \varsigma$, and thus the flux and $\gamma$-ray
luminosity, $F_\gamma$ and $L_\gamma$, are linearly proportional
to the cosmic ray enhancement. We compute secondary particle production
(electrons from knock-on interactions and electrons and positrons
from charged pion decay), and solve the loss equation with
ionization, synchrotron, bremsstrahlung, inverse Compton and
expansion losses (details are given in the papers \cite{18,20}).

We now compute $\gamma$-ray fluxes in a concrete example, and
following Section 1, we consider $\sim$2 M$_\odot$ of target mass
being modulated within $\sim$1 pc. The average density is $\sim25$
cm$^{-3}$. This amount of mass is typical of the configurations
studied in Section previously within the innermost $20\,R_c \sim 2-8$ pc.
To fix numerical values, we consider that the group of stars is at
a Galactic distance of 2 kpc. Using the computations of secondary
electrons and their distribution, we calculate the $\gamma$-ray flux
when the proton spectrum has a slope of 2.3 and 2.0. In the latter
case, to simplify, we show in Fig. \ref{f7}  only the pion decay contribution which
dominates at high energies,   produced by the whole cosmic ray
spectrum.

\begin{figure*}[t]
\centering
\includegraphics[width=5cm,height=6cm]{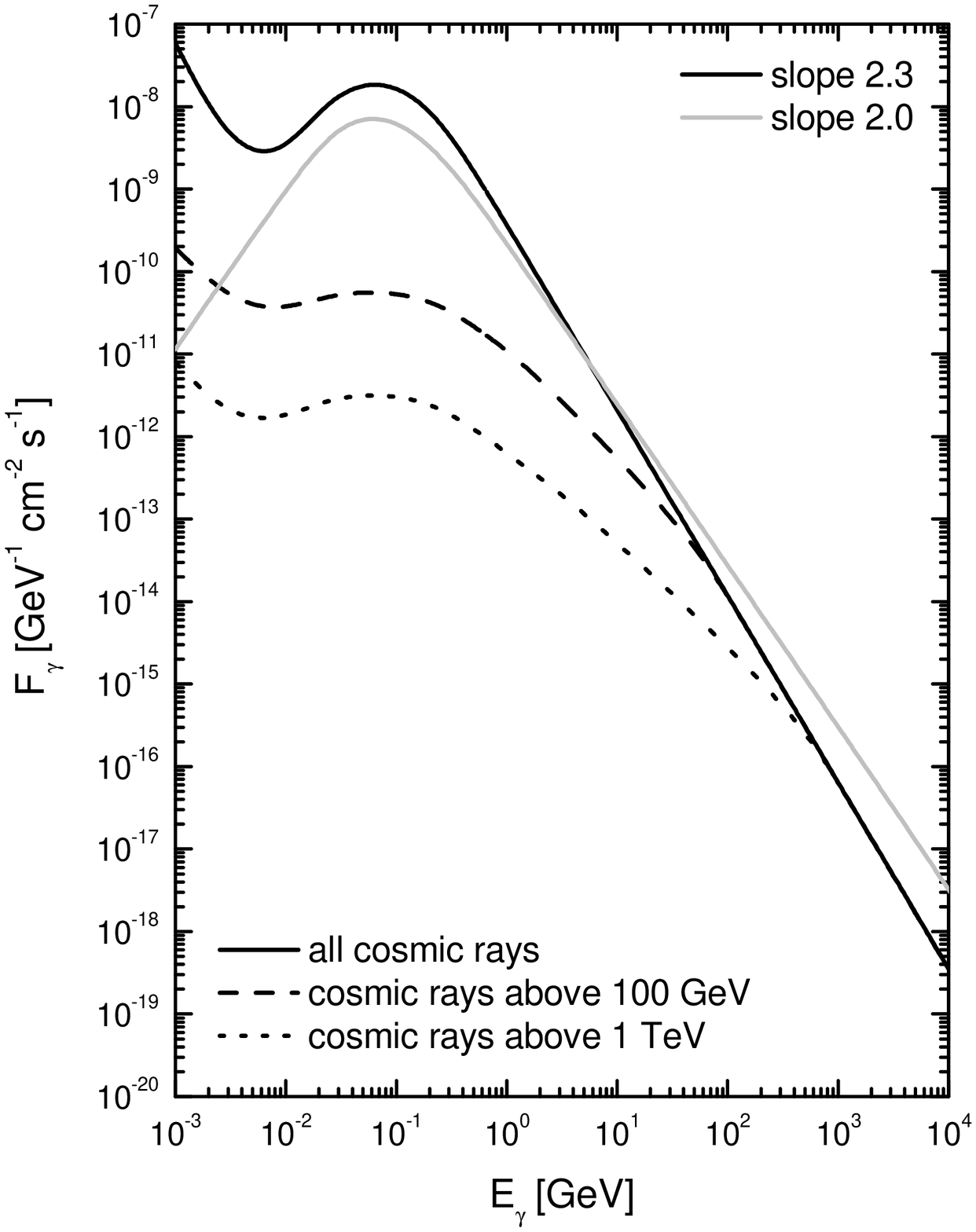}
\includegraphics[width=5cm,height=6cm]{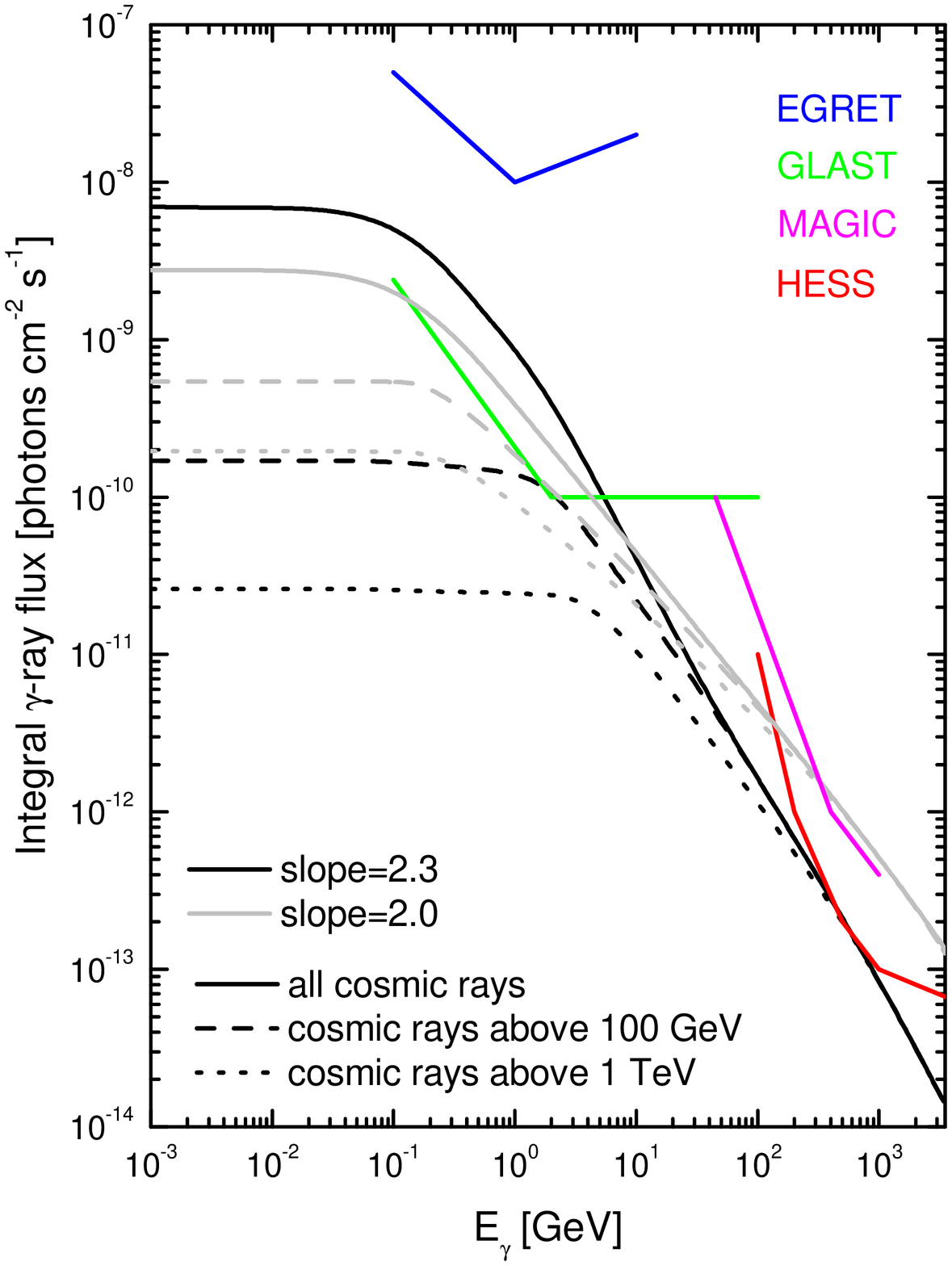}
\includegraphics[width=5cm,height=6cm]{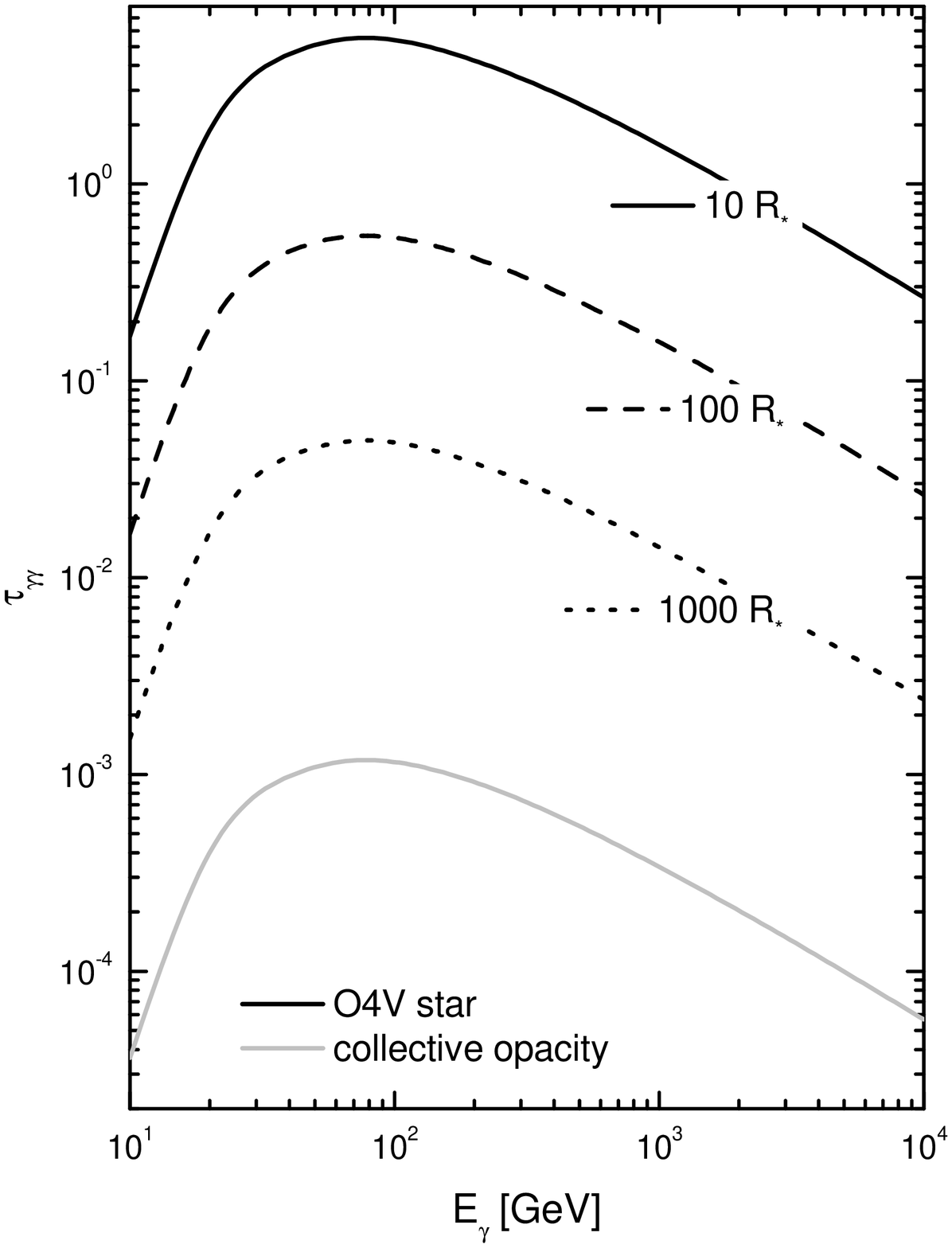}
\caption{Differential (left) and integral (right) fluxes of
$\gamma$-rays emitted in a non-modulated and a modulated
environment. The bump at very low energies in the left panel is
produced because we show leptonic emission coming only from
secondary electrons. Above $\sim 70$ MeV the emission is dominated
by neutral pion decay. Also shown are the EGRET, GLAST, MAGIC and
HESS sensitivities. Note that a source can be detectable by IACTs
and not by GLAST, or viceversa, depending on the slope of the
cosmic ray spectrum and degree of modulation. Right: Opacities to
$\gamma\gamma$ pair production in the soft
 photon field of an O4V-star at 10, 100 and 1000 $R_\star$, and in the collective photon field of
 an association with 30 stars distributed uniformly over a sphere
 of 0.5 pc. The closest star to the creation point is assumed to be
 at 0.16 pc, and the rest are placed following the average stellar density
 as follows: 1 additional star within
 0.1, 2 within 0.25, 4 within 0.32, 8 within 0.40 and 14
 within 0.5 pc.} \label{f7}
\end{figure*}

The differential photon flux is given by $ F_\gamma(E_\gamma) =[ {
V}/{4\pi D^2}] {Q_\gamma(E_\gamma)} = [ { M}/{m_p\,4\pi
D^2}][Q_\gamma(E_\gamma)/n] ,$  where $V$ and $D$ are the volume
and distance to the source, and $M$ the target mass. In those
examples where the volume, distance and/or the medium density are
such that the differential flux and the integral flux obtained
from it above 100 MeV with the full cosmic ray spectrum  is
greater than instrumental sensitivity, a modulated spectrum with a
100 GeV or a 1 TeV energy threshold might not produce a detectable
source in this energy range. However, the flux will be essentially
unaffected at higher energy. The left panel of Fig.
\ref{f7} shows that wind modulation can imply that a source may be
detectable for the ground-based Cerenkov telescopes without even being
close to be detected by instruments in the 100 MeV -- 10 GeV
regime (like  EGRET or the forthcoming GLAST). The right panel of
Fig.  \ref{f7} presents the integral flux of $\gamma$-rays as a
function of energy, together with the sensitivity of ground-based
and space-based $\gamma$-ray telescopes. The sensitivity curves
shown are for point-like sources; it is expected that extended
emission would require about a factor of 2 more flux to reach the
same level of detectability. Table 3 summarizes these results.
From Table 3 and Fig.  \ref{f7} we see that there are different
scenarios (possible relevant parameters are distance, enhancement,
degree of modulation of the cosmic ray spectrum and slope) for
which sources that shine enough for detection in the GLAST domain
may not do so in the IACTs energy range, and viceversa.

Finally, In Fig.  \ref{f7} (rightmost panel) we show the value of the photon photon opacity,
$\tau(E_\gamma)$, for different photon creation sites distant from
a O4V-star 10, 100, and 1000 $R_\odot$, with $R_\star = 12R_\odot$
and $T_{\rm eff} = 47400$ K. Unless a photon is created hovering
the star, well within 1000 $R_\star$, $\gamma$-ray opacities are
very low and can be safely neglected. This is still true for
associations in which the number of stars is some tens. Consider
for instance a group of 30 such stars within a region of 0.5 pc
(the central core of an association). The stellar density is given
by Eq. (\ref{n}); and the number of stars within a circle of
radius $R$ progresses as ${\cal N}= N (R/R_c)^3$. Fig.
 \ref{f7} shows the collective contribution to the opacity
 in this configuration is also very
low, since the large majority of the photons are produced far from
individual stars. However, this is not the case if one considers
the collective effect of a much larger association like the center of Cygnus OB
2 (\cite{21}). Reimer
demonstrated that even when a subgroup of stars like the ones
considered here is separated from a super cluster like Cygnus OB2
by about 10 pc, the influence of the latter produces an opacity
about one order of magnitude larger than that produced by the
local stars. But even in this case, Fig.   \ref{f7} 
shows that this opacity is not enough to preclude escape from the
region of the local enhancement of stellar density.
\section{Conclusions}

We have studied collective wind configurations produced by a
number of massive stars, and obtained densities and expansion
velocities of the stellar wind gas that is  target for hadronic
interactions in several examples. We have computed secondary
particle production, electrons and positrons from charged pion
decay, electrons from knock-on interactions, and solve the
appropriate diffusion-loss equation with ionization, synchrotron,
bremsstrahlung, inverse Compton and expansion losses to obtain
expected $\gamma$-ray emission from these regions, including in an
approximate way the effect of cosmic ray modulation. Examples
where different stellar configurations can produce sources for
GLAST and the MAGIC/HESS/VERITAS telescopes in non-uniform ways,
i.e., with or without the corresponding counterparts were shown.
 Cygnus OB 2 and Westerlund 1 maybe  two
associations where this scenario could be at work (\cite{20}).

\section*{Acknowledgments}

DFT has 
been supported by Ministerio de Educaci\'on y Ciencia (Spain) 
under grant AYA-2006-0530 and the Guggenheim Foundation.

\end{document}